# Competing magnetic states in a non-coplanar Kagome magnet

**Author Names**
Xiaodong Hu[1], Amar Fakhredine[2], Roger Guzman[3], Martin Frentrup[1], Jinan Shi[3], Giulio I. Lampronti[1], Sami El-Khatib[4], Wai Chuen Tse[5], Laura Gorzawski[1], Angelo Di Bernardo[6,7], Nadia Stelmashenko[1], Wu Zhou[3], Mehmet Egilmez[4,8], Danfeng Li[5], Grzegorz P. Mazur[9], Mario Cuoco[10], Carmine Autieri[2,10], Jason W. A. Robinson[1*]

**Author Affiliations**
[1] Department of Materials Science & Metallurgy, University of Cambridge, 27 Charles Babbage Road, Cambridge, CB3 0FS, United Kingdom.
[2] International Research Centre Magtop, Institute of Physics, Polish Academy of Sciences, Aleja Lotników 32/46, Warsaw, PL-02668, Poland.
[3] School of Physical Sciences, University of Chinese Academy of Sciences, Huairou District, Beijing, 100049, China.
[4] Department of Physics, American University of Sharjah, University City, Sharjah, 26666, United Arab Emirates.
[5] Department of Physics, City University of Hong Kong, 83 Tat Chee Ave, Kowloon Tong, Hong Kong, China.
[6] Department of Physics, University of Salerno, Via Giovanni Paolo II 132, Fisciano SA, 84084, Italy.
[7] Department of Physics, University of Konstanz, Universitätsstraße 10, Konstanz, 78464, Germany.
[8] Materials Research Center, American University of Sharjah, University City, Sharjah, 26666, United Arab Emirates.
[9] Department of Materials, University of Oxford, Parks Road, Oxford, OX1 3PH, United Kingdom.
[10] SPIN-CNR, UOS Salerno Fisciano (SA), IT-84084, Italy.

***Corresponding author name:** Jason W. A. Robinson
**Email:** jjr33@cam.ac.uk

**This PDF file includes:**

Main Text
Figures 1 to 4

**Abstract**

Understanding and controlling Berry-curvature-driven electronic transport is a central challenge in the field of spin-electronics. Antiferromagnetic (AFM) materials have potential applications in spin-electronic devices for memory, logic and data transfer. Whilst the majority of spin-electronic devices are based on ferromagnets, AFMs can offer greater magnetic robustness and faster dynamics without unwanted stray magnetic fields. However, conventional AFMs do not show large electrical signals for readout operation, which limits their applications. AFMs with a non-collinear Kagome lattice such as the $Mn_3X$ family (X = Sn, Ga, Ge, Ir, Pt) show a switchable anomalous Hall effect which could be exploited for both readout and writing operations within a device [Nature Reviews Materials 10(2), 109-127 (2025)]. However, their relatively strong in-plane anisotropy means it is hard to generate a non-coplanar spin structure which should lift the spin degrees of freedom for Berry curvature tuning. Here, we report an unconventional cubic phase of $Mn_3Ge$ with much lower in-plane anisotropy, enabling access to a non-coplanar spin structure. In the non-coplanar state, anomalous Hall measurements reveal magnetic-field-driven sign reversal and hump-like features consistent with competing magnetic states. Our findings establish a platform for non-coplanar magnetism in AFM spintronics.

**Significance Statement**

Antiferromagnets are attractive for high-density and high-speed spintronic devices due to the absence of stray magnetic fields and intrinsically fast spin dynamics, but their electrical readout is weak. Mn-based Kagome non-collinear antiferromagnets overcome this limitation by providing large electrical signals associated with their spin structures. We show that a cubic $Mn_3Ge$ phase hosts an intrinsic non-coplanar spin structure, expanding the functionality of Kagome antiferromagnetic spintronics. In this system, the anomalous Hall response exhibits magnetic-field-driven sign reversal and hump-like features up to 400 K. Our measurements and calculations are consistent with a phenomenological model in which magnetic-field sweeps drive the system through competing non-coplanar magnetic configurations. This establishes cubic $Mn_3Ge$ as a platform for controlling Berry-curvature-driven electronic transport through non-coplanar magnetism.

**Main Text**

**Introduction**

Antiferromagnetic (AFM) materials are promising for spin-electronic (spintronic) applications because of the absence of a net magnetic moment in zero magnetic field, robustness against magnetic perturbations, and fast spin dynamics [1, 2]. However, in contrast to ferromagnetic (FM) materials, AFMs generally show weak electrical readout signals due to the cancellation of magnetotransport contributions. Therefore, identifying AFMs that can generate large magnetotransport effects is important for the development of AFM spintronics. Among them, non-collinear Kagome AFMs $Mn_3X$ (X = Sn, Ga, Ge, Ir, Pt) are of particular interest because their chiral spin textures break time-reversal symmetry and, when combined with spin-orbit coupling (SOC), generate a finite Berry curvature in momentum space and a large anomalous Hall effect (AHE) [3-13]. Electrical switching of non-collinear AFM order via spin-orbit torque has been demonstrated in $Mn_3X$, with tunnelling magnetoresistance values up to 100%, establishing a framework for AFM spintronics [14, 15].

While previous studies have focused on the coplanar non-collinear spin states of Kagome $Mn_3X$ [16], non-coplanar spin textures and their associated magnetotransport properties are less well understood. One reason is the strong in-plane anisotropy in $Mn_3X$ requires a large magnetic field to cant the spins out of the Kagome plane. To the best of our knowledges, non-coplanar spin states in Kagome AFMs have only been studied in the spin glass state (SGS) of $Mn_3Sn$ below 50 K [17, 18] or via interfacial spin-orbit coupling (SOC) at Pt/$Mn_3Sn$ interfaces [19, 20]. Here, we report thin films of cubic-phase $Mn_3Ge$ with an intrinsic non-coplanar spin texture. Our calculations suggest that the coexistence of symmetric and antisymmetric exchange interactions can support competing stable and metastable magnetic states within a single structural phase. The observed unconventional AHE with a magnetic-field-induced sign reversal and a hump-like feature, which are absent in coplanar Kagome AFMs, are consistent with a phenomenological model involving field-driven transitions between these competing magnetic states.

## Results

### A. Growth of $Mn_3Ge$ thin films

We deposited textured $Mn_3Ge$ thin films onto Ru-buffered $Al_2O_3(0001)$ substrates. In bulk material, hexagonal-phase $Mn_3Ge$ (h-$Mn_3Ge$) is thermodynamically stable with space group $P6_3/mmc$ (No. 194), while cubic-phase $Mn_3Ge$ (c-$Mn_3Ge$) has a centrosymmetric structure with space group $Pm\bar{3}m$ (No. 221) [21]. Out-of-plane XRD $2\theta$-$\omega$ scans, collected in a coupled geometry with an incident angle offset of 0.2° in $\omega$, show the $(0001)$ h-$Mn_3Ge$ and $(111)$ c-$Mn_3Ge$ reflections. Additional scans over a large range of 30° ≤ $2\theta$ ≤ 110° do not show any other detectable phases [Fig. 1(a)]. RSMs comparing a 100-nm-thick $Mn_3Ge$ film grown on Ru/$Al_2O_3$ [Fig. 1(b), top] with a Ru/$Al_2O_3$ reference sample [Fig. 1(b), bottom] show the $(10\bar{1}3)$ Ru and $(11\bar{2}9)$ $Al_2O_3$ reflections and that the Ru and $Al_2O_3$ are 30° rotated to another. This rotation corresponds to the honeycomb epitaxial growth of Ru on $Al_2O_3(0001)$ [22], providing a template on which h-$Mn_3Ge$ grows by matching a doubled in-plane Ru unit cell without rotation. The in-plane epitaxial relationship between the hexagonal and cubic phases was characterized by pole-figure measurements [Fig. 1(c)]. Both the $\{20\bar{2}3\}$ h-$Mn_3Ge$ and $\{111\}$ c-$Mn_3Ge$ reflections have matching patterns with sixfold symmetry due to similar atomic arrangements in their Kagome planes.

STEM-HADDF imaging [Fig. 1(d), left] shows a sharp and continuous interface, and confirms the epitaxial relationship $Al_2O_3(0001)[2\bar{1}\bar{1}0]$ || $Ru(0001)[10\bar{1}0]$ || h-$Mn_3Ge(0001)[10\bar{1}0]$ || c-$Mn_3Ge(111)[\bar{1}10]$. The enlarged region [Fig. 1(d), right] aligned with the atomic models reveals a structural transition within the $Mn_3Ge$ layer: the first 3-5 nm adjacent to the Ru buffer exhibits the hexagonal phase, above which the film converts to the cubic phase. Fast Fourier transform (FFT) patterns of c-$Mn_3Ge$ and h-$Mn_3Ge$ regions are consistent with the expected diffraction patterns of the two phases [Fig. 1(e)]. The superlattice reflections in c-$Mn_3Ge$ regions arise from the ordered repetition of Mn and Ge atomic arrays, indicating high crystalline quality. Atomic models of the two phases, highlighting the resolved structural motifs, are shown in Fig. 1(f).

### B. AHE with sign reversal and hump-like features

We measured magnetotransport properties of a 100-nm-thick $Mn_3Ge$ film in a Hall-bar geometry [Fig. 2(a)], varying the temperature and maximum field sweep range ($\Delta H^*$). The magnetic field was applied perpendicular to the film plane, along the h-$Mn_3Ge$ $[0001]$ and c-$Mn_3Ge$ $[111]$ directions. The zero-field longitudinal resistivity ($\rho_{xx}$) varies smoothly with temperature from 2 K to 300 K without any low-temperature upturn or kink, indicating the absence of a SGS transition [Fig. 2(b)]. We calculated the Hall conductivity ($\sigma_{xy}$) from the measured Hall voltage by removing the ordinary Hall effect (OHE) and normalizing by the film thickness and longitudinal resistivity [Figs. 2(c-j)]. This normalization is shown in Fig. S2 in the Supplemental Material [23], which allows a direct comparison of $\sigma_{xy}$ across films of different thicknesses and at different temperatures.

We extract the Hall conductivity remanence ($\sigma_{xy}^R$) from the hysteresis loops, defined as the difference in $\sigma_{xy}$ at zero magnetic field after sweeping the field down from positive ($\sigma_{xy}^R(+)$) and sweeping up from negative ($\sigma_{xy}^R(-)$) values [Figs. 2(c,f)]. For small $\Delta H^*$ [Figs. 2(c,g)], the shape of the loops matches a conventional FM with positive $\sigma_{xy}^R$ and positive $\sigma_{xy}$ at maximum magnetic field. As $\Delta H^*$ increases [Figs. 2(d-f,h-j)], $\sigma_{xy}^R$ and $\sigma_{xy}$ at maximum field reverse in sign from positive to negative, and the loops develop flat regions and hump-like features. For example, in Fig. 2(j), as the field is swept from +14 T to –14 T, $\sigma_{xy}$ remains approximately constant until about 0 T in the flat region, then decreases to form a hump with a minimum near –3 T. On the return sweep (–14 T to +14 T), $\sigma_{xy}$ stays flat until about 0 T and then increases to a maximum around +3 T.

We quantify the hump amplitude by $\sigma_{xy}^{hump} = [|\sigma_{xy}^{hump}(+)| + |\sigma_{xy}^{hump}(-)|]/2$. Here, $\sigma_{xy}^{hump}$ (+) is the difference between the peak $\sigma_{xy}$ and the flat-region $\sigma_{xy}$ on the positive-field branch, and $\sigma_{xy}^{hump}$ (–) is the corresponding difference on the negative branch [Figs. 2(d,j)]. At low $\Delta H^*$, when the peak $\sigma_{xy}$ equals that of the flat region, the hump amplitude vanishes [Figs. 2(c,g)].

Transport measurements on $Mn_3Ge$ films with thicknesses less than 5 nm, where the structure is predominantly hexagonal, are shown in Fig. S3 in the Supplemental Material [23]. After subtracting the OHE, the Hall signal is negligible compared with that in the cubic phase, consistent with previous studies [7-9, 24]. Hence, the sign reversal and the hump feature are related to c-$Mn_3Ge$.

**C. Evolution of sign reversal and hump in AHE with temperature and magnetic field**

We mapped the evolution of $\sigma_{xy}^R$ and $\sigma_{xy}^{hump}$ for $\Delta H^*$ = 1-14 T over a temperature range of 2-400 K. As shown in Fig. 3(a), $\sigma_{xy}^R$ increases with $\Delta H^*$ at small $\Delta H^*$, then decreases and reverses in sign at large $\Delta H^*$. In contrast, the magnetization remanence $M^R$ [Fig. 3(b)], defined similarly to $\sigma_{xy}^R$, grows monotonically with $\Delta H^*$. The extracted 2 K and 300 K trends [Fig. 3(c)] highlight this contrast: $\sigma_{xy}^R$ increases with $\Delta H^*$ up to around 4 T, then decreases and reverses in sign between $\Delta H^*$ of 5 T and 9 T depending on temperature, whereas $M^R$ remains positive and grows monotonically with $\Delta H^*$. Thus, the sign reversal and hump feature cannot be accounted for by a magnetization-driven AHE alone.

In hysteresis loops measured over a field range of $\Delta H^*$ = 10 T [Figs. 2(e,i)], $\sigma_{xy}^{hump}$ at 300 K is larger than that at 2 K, and $\sigma_{xy}^R$ reverses in sign between the two temperatures. This indicates that the $\Delta H^*$ values required for the hump emergence and the sign reversal depend on temperature. As shown in Fig. 3(d), the hump emerges at $\Delta H^*$ around 5 T and $\sigma_{xy}^R$ reverses sign near 9 T at 2 K, whereas at 300 K these features occur at approximately 3 T and 5.5 T, respectively. Therefore, the magnetic field required to trigger these anomalous behaviors decreases as the temperature increases. The critical $\Delta H^*$ for the sign reversal in $\sigma_{xy}^R$ is always larger than that for the hump emergence.

**D. Possible mechanisms**

The results highlight two key features in the behavior of $\sigma_{xy}$ depending on the field sweep range and temperature: (1) an enhancement of $\sigma_{xy}$, which coincides with a hump-like feature; (2) a sign reversal of $\sigma_{xy}^R$ and $\sigma_{xy}$ at the highest applied field. In general, hump-like features in Hall measurements can be attributed to a topological Hall effect (THE) from real-space Berry curvature [17-20, 25], or to the superposition of opposing AHE channels from distinct structural or magnetic regions [26-28]. In principle, both mechanisms are plausible for c-$Mn_3Ge$. Out-of-plane spin canting in a Kagome spin arrangement could produce a finite solid angle, leading to nonzero scalar spin chirality and real-space Berry curvature. Alternatively, the coexistence of h-$Mn_3Ge$, c-$Mn_3Ge$, and their interface could give rise to multiple AHE contributions with opposite signs.

To investigate a possible THE origin, we measured minor loops of $\sigma_{xy}$, sweeping the magnetic field from 9 T to a maximum negative value $H_{n,max}$ and then back to 9 T [Fig. 3(e)]. The positive field hump appears only when $H_{n,max}$ extends beyond approximately –3.5 T, where the negative field hump develops. Therefore, the positive field Hall response depends on the negative field history. This hysteretic minor-loop behavior has previously been shown to be inconsistent with a THE scenario [29-31]. This is supported by a study in which skyrmions were independently confirmed by magnetic imaging, and the genuine THE showed no hysteretic dependence on $H_{n,max}$ [32]. In that system, the THE was also limited in a temperature range and showed no temperature driven sign reversal of $\sigma_{xy}$, whereas in our c-$Mn_3Ge$, the hump persists from 2 to 400 K and $\sigma_{xy}$ reverses sign with temperature. Together, these observations argue against a simple scalar-chirality-driven THE in which the chiral texture is determined by the instantaneous magnetic field. However, transport alone cannot exclude more complex chiral textures that themselves exhibit hysteresis and field-history dependence [33].

Although the AHE from the (0001)-oriented h-$Mn_3Ge$ is negligible [7, 8, 23, 24], opposing AHE contributions from c-$Mn_3Ge$ and the hexagonal–cubic interface could in principle produce a hump [34]. However, our findings show that $\sigma_{xy}^{hump}$ increases with film thickness from 30 to 100 nm and then saturates [Fig. 3(f)]. This is inconsistent with a simple two-AHE picture, in which $\sigma_{xy}^{hump}$ would decrease with increasing film thickness [28], as the AHE contribution from c-$Mn_3Ge$ progressively outweighs a fixed interfacial contribution. Interfacial effects could nevertheless be more complex, for example through strain or magnetic state nucleation. Therefore, we systematically investigated the thickness-dependent lattice parameters by XRD, and found that the c-$Mn_3Ge$ peak positions remain approximately unchanged for films with thicknesses of 30 to 200 nm [Fig. S1(b)], indicating no pronounced strain relaxation. We subsequently optimized c-$Mn_3Ge$ growth on Ru-buffered SiC, where no h-$Mn_3Ge$ phase is detectable by XRD [Fig. S1(c)], but the transport features including the hump, sign reversal of $\sigma_{xy}$, and $\Delta H^*$-dependence remain, as shown in Fig. S4. These results indicate that a detectable hexagonal–cubic interface is not required for the observed Hall features, although subtle interfacial effects on magnetic-state nucleation cannot be excluded.

More broadly, field-history-dependent magnetic and transport responses are found in other magnetic systems with spin-glass phases [35], skyrmions [36-39], and distinct spin populations [40]. In c-$Mn_3Ge$, a spin-glass phase is inconsistent with the absence of a low-temperature spin-glass transition, and a simple skyrmion-related THE is unlikely given the minor-loop and temperature dependency discussed above. This leaves multiple-AHE channels associated with distinct spin populations as another alternative interpretation. In the Kagome ferromagnet $Co_3Sn_2S_2$, a similar $\Delta H^*$-dependent Hall response was attributed to a secondary spin population with a larger coercive field than the majority population, supported by spatially varying local Hall measurements [40], and a separate NMR study showing Co-site equivalence breaking above 90 K [41].

For c-$Mn_3Ge$, distinct spin populations could arise in two scenarios. First, Mn sites within a single Kagome lattice may become magnetically inequivalent. However, the AHE in Kagome AFMs is determined by the Berry curvature of the full magnetic configuration rather than by the net magnetization alone. For example, different compensated in-plane spin arrangements can exhibit opposite AHEs [11, 21, 42]. Therefore, canting or site inequivalence can modify the electronic structure and $\sigma_{xy}$, as shown in Fig. S7, but cannot be separated into independent magnetization-driven AHE contributions. This is consistent with the different $\Delta H^*$-dependences of $\sigma_{xy}^R$ and $M^R$ in Figs. 3(a,b), showing that the AHE need not follow the net magnetization as in conventional ferromagnets. Second, spatially distinct majority and minority Mn populations with different coercive fields, analogous to $Co_3Sn_2S_2$, could arise from off-stoichiometry or chemical disorder and produce a $\Delta H^*$-dependent Hall response. Our local electron energy loss spectroscopy (EELS) showed no pronounced Mn enrichment [Fig. S1(a)], consistent with the $Mn_{75}Ge_{25}$ composition reported for c-$Mn_3Ge$ [21]. Additionally, the superlattice reflections indicate good chemical ordering [Fig. 1(e)], and similar local Hall measurements show no strong spatial variation in the AHE on length scales of 5 μm, as shown in Fig. S5. These observations make a substantial spatially inhomogeneous minority-spin population less likely, although inhomogeneity at shorter length scales cannot be excluded.

**E. Exchange interactions and magnetic states in c-$Mn_3Ge$**

To construct a phenomenological interpretation of the hump-like feature and sign reversal, we investigated the magnetic energy landscape of c-Mn3Ge using density functional theory (DFT) calculations. In non-collinear Kagome AFMs, vector spin chirality characterizes the magnetic ground state by describing the spin rotation in a triangular arrangement of three Mn atoms. When moving anticlockwise from one Mn atom to the next, the spins rotate anticlockwise for a chirality of +1 and clockwise for a chirality of –1. The ground-state chirality of c-$Mn_3Ge$ is +1 [21], consisting of two degenerate spin configurations A and B, as shown in Figs. 4(a,b). The two configurations may coexist in different magnetic domains and generate opposite signs of $\sigma_{xy}$ through the momentum-space Berry curvature [43, 44]. This Berry curvature is influenced by the Weyl points (connected by inversion symmetry) and semi-Dirac points near the Fermi level in the band structure [21, 23].

We define the $z$ axis along the $[111]$ direction of c-$Mn_3Ge$ (along which the magnetic field is applied in magnetotransport measurements). Within relativistic DFT calculations, c-$Mn_3Ge$ can exhibit a magnetic solution with vanishing magnetization along $z$, consistent with previous studies [8, 21, 45]. However, by uniformly increasing the initial values of the spin component along $z$ ($S_z$) on all Mn atoms, the system converges to FM solutions with finite magnetization. This indicates the presence of a symmetric exchange interaction in c-$Mn_3Ge$ that induces FM coupling for $S_z$. In Kagome AFMs, the dominant exchange interaction between Mn atoms is antiferromagnetic among the in-plane spin components, stabilizing an all-in- or all-out-type ground state [46]. The persistence of the AHE up to 400 K indicates that the in-plane AFM interactions remain robust despite the geometric frustration of the Kagome lattice. This would be surprising for h-$Mn_3Ge$, which has a Néel temperature ($T_N$) of 365–400 K [7–9, 24], whereas c-$Mn_3Ge$ has a $T_N$ of up to 490 K [21], consistent with our observation.

To account for the two distinct exchange interactions, we describe the system using an XXZ Hamiltonian [47]:

$$H = \sum_{i<j}\left[J_{ij}^{\parallel}\left(S_i^x S_j^x + S_i^y S_j^y\right) + J_{ij}^{\perp} S_i^z S_j^z\right],$$

where $S_i^\alpha$ denotes the spin component at site $i$ along direction $\alpha$, $J^{\parallel} > 0$ is the strong AFM in-plane exchange coupling, and $J^{\perp} < 0$ is the FM out-of-plane exchange coupling. We note that a two-up/one-down out-of-plane canting configuration has been proposed for $Mn_3Sn$ at low temperature [17]. A subsequent study identified this magnetic phase as a spin-glass state below 50 K [18]. However, this spin configuration is higher in energy than the uniform canting configuration by $4|J_\perp|S_z^2$ per Kagome lattice because $J_\perp < 0$. Additionally, no SGS transition was observed in our $\rho_{xx}$-$T$ measurements [Fig. 2(b)]. It was shown that the first neighbor interaction dominates in intermetallic compounds with Kagome lattices [48]. The opposite signs of the first-neighbor $J^{\parallel}$ and $J^{\perp}$ suggest an indirect Ruderman-Kittel-Kasuya-Yosida (RKKY) exchange mechanism mediated by conduction electrons. The RKKY exchange is known to oscillate in sign depending on the Fermi surface, which allows different signs for $J^{\parallel}$ and $J^{\perp}$ [49]. Consequently, the coexistence of the in-plane AFM coupling ($J^{\parallel} > 0$) and the out-of-plane FM coupling ($J^{\perp} < 0$) in c-$Mn_3Ge$ arises within an XXZ description governed by RKKY exchange interaction.

The symmetric FM exchange interaction produces an even double-well energy profile as a function of $S_z$, which is identical for spin configurations A and B and exhibits two degenerate minima at opposite canting directions. In contrast, due to SOC and broken PT symmetry, the antisymmetric Dzyaloshinskii-Moriya interaction (DMI) induces a weak ferromagnetism via spin canting out of the Kagome plane and favors one canting direction over the other. This preferred canting direction is reversed between spin configurations A and B. Therefore, the DMI contributes an energy term that is odd in $S_z$ [50], and the sign of this energy contribution is opposite for the two spin configurations. As detailed in Fig. S8 in the Supplemental Material [23], the coexistence of symmetric and antisymmetric exchange interactions gives rise to the energy profiles shown in Figs. 4(c,d) as functions of $S_z$. For each spin configuration, two local energy minima are related by a mirror reflection of spin canting with respect to the Kagome plane. Since mirror symmetry is absent in the system, reflecting the spins across the Kagome plane produces a stable ground state (S) and a higher-energy metastable state (M). These minima are shifted in opposite canting directions for the two configurations: configuration A has a stable state with $S_z > 0$ and a metastable state with $S_z < 0$, whereas configuration B shows the opposite. Both exchange interactions act along the $z$ direction, while the net magnetization within the Kagome plane remains zero.

**F. Phenomenological model of field-driven transitions between magnetic states**

We interpret the hump-like features and AHE sign reversal by a phenomenological model based on field-driven transitions between the stable and metastable states in two spin configurations. The spin schematics in Fig. 4(e) of the AHE over $\Delta H^* = 14$ T indicate the dominant spin configuration contributing to the AHE in the numbered regimes. At positive saturation (regime 1, +14 T), the contribution to the AHE is dominated by the stable ground state with spins canting up (stable A state), corresponding to a negative $\sigma_{xy}$. This contribution remains dominant as the magnetic field is swept towards 0 T (regimes 1 to 2, +14 to 0 T), producing the flat region. As the magnetic field is swept from positive to negative values, two competing processes occur simultaneously:

{1} the stable A state converts to the metastable A state with spins canting down;

{2} the metastable A state converts to the stable B state with in-plane spin reorientation.

Since process {2} requires additional energy for the in-plane spin reorientation vs process {1}, process {1} and contribution from metastable A state dominate at low field (regimes 2 to 3, 0 to -3 T). The larger $|\sigma_{xy}|$ around regime 3 than that around regime 1 indicates an enhanced $\sigma_{xy}$ in the metastable state vs the stable state, consistent with theoretical predictions [51, 52]. Upon further increase of the magnetic field (regimes 3 to 4, -3 to -14 T), process {2} and contribution from stable B state dominate, with $\sigma_{xy}$ of opposite sign to configuration A, and the net value of $\sigma_{xy}$ approaches 0 before reversing in sign. The transitions among the magnetic states during the magnetic field sweep back are similar. Stable B state contribution remains dominant until about 0 T (regimes 4 to 5, -14 to 0 T) in the flat region, and then two competing processes occur as the field is swept from negative to positive values: {1} the stable B state converts to the metastable B state, and {2} the metastable B state converts to the stable A state. The contribution from metastable B state dominates at low field (regimes 5 to 6, 0 to +3 T) with a larger $|\sigma_{xy}|$ than the stable B state, causing the increase in $\sigma_{xy}$. The contribution from stable A state dominates at high field (regimes 6 to 1, +3 to +14 T) with $\sigma_{xy}$ opposite in sign to configuration B, and the net value of $\sigma_{xy}$ approaches 0 before reversing in sign. We note that different spin configurations can exhibit different $|\sigma_{xy}|$, and thus dominant contribution to the AHE does not necessarily correspond to the most populated magnetic state or the state dominating the net magnetization.

More generally, within this phenomenological model, the emergence of the hump and the sign of $\sigma_{xy}$ at maximum field depends on a competition between two processes as the field is swept:

{1} the conversion of the stable state into the metastable state with the same in-plane spin configuration, which increases $|\sigma_{xy}|$;

{2} the conversion of the metastable state into the stable state with a different in-plane spin configuration, which decreases $|\sigma_{xy}|$ and then reverses its sign.

Process {1} and the contribution from metastable states dominate at low magnetic field, while process {2} and the contribution from stable states dominate at high field. The relative influence of processes {1} and {2} determines the shape of the hysteresis loop. For small $\Delta H^*$ [Figs. 2(c,g)], process {1} and the contribution from metastable states dominate during the entire loop. Therefore, $\sigma_{xy}$ varies without a hump and the $\sigma_{xy}$ at the maximum field is determined by the metastable states (positive at positive field). For intermediate $\Delta H^*$ [Figs. 2(d,h)], there is a critical field where the rate of increase in $|\sigma_{xy}|$ from process {1} equals to the rate of decrease in $|\sigma_{xy}|$ from process {2}. At this point, the net value of $|\sigma_{xy}|$ reaches a maximum and forms a hump before decreasing. Therefore, the hump appears once $\Delta H^*$ is high enough that both processes occur at the same rate. Upon further increase in $\Delta H^*$ [Figs. 2(f,j)], $\sigma_{xy}$ contribution from the stable state cancels out that from the metastable state, and $\sigma_{xy}$ at the maximum field in the loop goes to 0 and reverses in sign.

We stress that the transport measurements do not determine the exact microscopic spin configuration in each numbered regime or the detailed transition pathway between them. The evolution of the in-plane spin configuration need not be a simple switch between two discrete spin configurations A and B. The reduced in-plane anisotropy of c-$Mn_3X$ [1] allows continuous in-plane spin rotation and coexistence of the magnetic configurations in the numbered regimes in Fig. 4(e), which can broaden the hump over a wider field range than in other materials [17–20, 25–29]. Direct identification of these spin configurations and their dynamics requires microscopic magnetic probes such as neutron scattering, which are challenging for AFM thin films and are left for future study. Nevertheless, out-of-plane canting modifies the relative stability of the in-plane spin configurations [Figs. 4(c,d)], allowing field-driven transitions between the stable and metastable states in our phenomenological model. This in-plane reorientation also explains why the metastable state dominates at lower fields even though the stable state is lower in energy. For example, during the positive-to-negative field sweep (regimes 1–4), the stable A state first evolves into the metastable A state by out-of-plane canting reversal at low negative fields (regimes 2–3), while reaching the stable B state requires overcoming an additional energy barrier for in-plane spin reorientation and occurs only at larger negative fields (regimes 3–4). At low temperature, reduced thermal activation allows the metastable state to persist over a wider field range, while increasing temperature facilitates the transition to the stable state. This is consistent with our observation that the $\Delta H^*$ required for the hump emergence and AHE sign reversal decreases with increasing temperature [Fig. 3(d)] and with reports of thermally assisted spin evolution in another Kagome system [53].

In summary, within our phenomenological model, the evolution of the Hall loops with $\Delta H^*$ can be understood by the relative AHE contributions from the stable and metastable states. At small $\Delta H^*$, even at the highest field, the AHE remains dominated by the metastable-state contribution, generating a hysteresis loop with no hump and positive $\sigma_{xy}^R$. At intermediate $\Delta H^*$, the $\sigma_{xy}$ contribution from the stable state starts to dominate but has not yet fully cancelled that from the metastable state, producing a hump in the loop and $\sigma_{xy}^R$ remains positive. At large $\Delta H^*$, the AHE evolves from metastable-state-dominated to stable-state-dominated, generating a hump and a negative $\sigma_{xy}^R$. Therefore, different field-sweep ranges produce distinct loop shapes because the dominant contribution to the AHE varies between the stable and metastable states. Overall, this $\Delta H^*$-dependent evolution is consistent with our phenomenological model under a soft magnetic response and in the absence of a spin-glass state.

Unlike c-$Mn_3Ge$, h-$Mn_3Ge$ lacks the combination of FM interaction, DMI, and associated $S_z$ canting that gives rise to two energy minima. Consequently, within the present model, h-$Mn_3Ge$ is expected to exhibit neither a metastable magnetic state nor an A-B spin configuration switching under a magnetic field. This is consistent with the absence of a hump or $\sigma_{xy}^R$ sign reversal in our h-Mn3Ge measurements and previous studies [7-9, 23, 24].

**Discussion and Summary**

We have demonstrated that cubic-phase Kagome $Mn_3Ge$ exhibits an AHE with a sign reversal and a hump-like feature in $\sigma_{xy}$ up to 400 K. These phenomena are consistent with a non-coplanar spin configuration emerging from the non-collinear AFM state. Our theoretical analysis suggests that this spin texture may result from the coexistence of conventional ferromagnetism induced by a symmetric RKKY exchange and weak ferromagnetism induced by an antisymmetric DMI. This coexistence generates an unconventional energy profile with stable and metastable magnetic states. We interpret the observed AHE features by a phenomenological model based on field-driven transitions between these competing magnetic states. Compared with a recent work on c-$Mn_3Ge$ [21], our films exhibit stronger magnetic and Hall responses. This difference may arise from different growth conditions and degree of chemical ordering, as indicated by the more pronounced superlattice reflections in our films. Chemical ordering could affect the magnetic and transport properties in the Kagome $Mn_3X$ systems since their magnetic and electronic structures are closely linked to the arrangement of Mn atoms in the Kagome lattice.

Hump-like features in the Hall effect, with or without a sign reversal, have previously been attributed either to real-space Berry-curvature effects such as skyrmions [54-56] and other engineered systems [17-20], often enabled by heavy-element-induced SOC. Hump-like features can also originate from multiple AHE channels, arising from different structural phases within the same film [26-28]. Our results suggest an alternative interpretation of hump-like and sign-reversal features in a single phase driven by a competition between magnetic states and the resulting changes in the momentum-space Berry curvature that underlie the AHE.

These results highlight the sensitivity of the electronic band structure to the spin texture and provide a platform to engineer momentum-space Berry curvature using an applied magnetic field in AFMs. In this way, cubic $Mn_3Ge$ offers a versatile system for exploring non-coplanar AFM spintronics. Further studies of the spin dynamics associated with these state transitions will be important for a complete understanding of the underlying mechanisms.

## Materials and Methods

### A. Sample preparation

The thin films were deposited by magnetron sputtering in an ultra-high-vacuum system with a base pressure below 1 × $10^{-8}$ Torr. $Al_2O_3(0001)$ or Si-terminated 4H SiC$(0001)$ substrates were loaded through a load-lock system without any additional surface preparation and were rotated at 10 rpm during deposition to ensure uniform film growth. The growth temperature was controlled by a radiative heater mounted on the back of the sample holder. The Ru buffer layer was deposited at 700°C by a radiofrequency (r.f.) power of 30 W in an Ar pressure of 1.5 mTorr. The $Mn_3Ge$ layer was subsequently deposited from a $Mn_3Ge$ alloy target at 700°C by a direct-current (d.c.) power of 30 W in an Ar pressure of 2 mTorr. The Hall bar geometries for transport measurements were fabricated by optical lithography and Ar ion-beam etching.

### B. Structural characterization

X-ray diffraction (XRD) was carried out using a four-circle Panalytical Empyrean vertical $\omega/\theta$ diffractometer equipped with a hybrid two-bounce primary monochromator. A proportional counter was used for the $2\theta$-$\omega$ and $\phi$ scans, and a two-dimensional PIXCEL position-sensitive detector (PSD) was used to collect reciprocal space maps (RSMs).

Cross-sectional samples for scanning transmission electron microscopy (STEM) were prepared by standard lift-out process using a focused ion beam (FIB) Thermo Fischer Scientific Helios G4 CX Dual-Beam. To minimize sidewall damage and create sufficiently thin samples, final milling was carried out at low voltages (2 kV).

Atomic-resolution STEM high-angle annular dark-field (HAADF) images were carried out in a JEOL GrandARM-2 operated at 200 kV. The annular detector collection semi-angles were set to 68-280 mrad for HAADF. EELS experiments were carried out in a NION U-HERMES microscope operated at 60 kV, equipped with a fifth-order aberration corrector, an alpha-type monochromator and the Dectris ELA direct electron detector for EELS acquisition. The experiments were conducted with beam convergence semi-angle and EELS collection semi-angle of 32 and 75 mrad, respectively, and an energy dispersion of 0.9 eV/ch and a pixel dwell time of 100 ms.

### C. Transport and magnetization measurements

Longitudinal and transverse resistances were measured using the fabricated Hall bar geometries by a four-probe configuration in a Quantum Design Physical Property Measurement System (PPMS), with a Keithley 6221 current source and a Keithley 2182A nanovoltmeter. Magnetic measurements were performed using a Quantum Design Magnetic Property Measurement System (MPMS3) with a vibrating sample magnetometer (VSM) and a superconducting quantum interference device (SQUID).

### D. Density functional theory calculation

Density functional theory (DFT) calculations were performed using the Vienna *ab initio* Simulation Package (VASP) [57, 58] with the projector-augmented wave (PAW) method [59]. The exchange-correlation functional was described using the generalized gradient approximation (GGA) in the Perdew-Burke-Ernzerhof (PBE) form [60]. Periodic boundary conditions were applied, and the Bloch theorem was used to treat the crystalline system. The plane-wave cutoff energy was set to 390 eV, and the total energy convergence criterion was $10^{-4}$ eV. The Brillouin zone was sampled using a Monkhorst-Pack 16 × 16 × 12 *k*-point mesh. To estimate the weak ferromagnetism, we used a reconstructed supercell oriented along the [111] direction of the cubic cell containing three formula units. The PAW potentials included 7 valence electrons for Mn and 4 for Ge, resulting in a total of 75 valence electrons for the 9 Mn and 3 Ge atoms in the supercell. The lattice parameters were taken from experiment as $a = b = 5.164$ Å and $c = 6.324$ Å. Maximally localized Wannier functions were constructed using the Mn 3*d* and Ge 4*p* orbitals as initial projections. These projections were used to obtain an accurate tight-binding Hamiltonian describing the low-energy electronic structure of the system.

**Acknowledgments**

R.G. was supported by the CAS Project for Young Scientists in Basic Research (YSBR-003), the National Natural Science Foundation of China (grant No. 52373231), and the Beijing Outstanding Young Scientist Program (No. BJJWZYJH01201914430039). This research benefited from resources and support from the Electron Microscopy Center at the University of Chinese Academy of Sciences. C.A. was supported by the "MagTop" project (FENG.02.01-IP.05-0028/23) carried out within the "International Research Agendas" programme of the Foundation for Polish Science, co-financed by the European Union under the European Funds for Smart Economy 2021-2027 (FENG). C.A. and M.C. acknowledge support from PNRR MUR project PE0000023-NQSTI. We further acknowledge access to the computing facilities of the Interdisciplinary Center of Modeling at the University of Warsaw, Grant g91-1418, g91-1419, g96-1808 and g96-1809 for the availability of high-performance computing resources and support. We acknowledge the access to the computing facilities of the Poznan Supercomputing and Networking Center, Grants No. pl0267-01, pl0365-01 and pl0471-01.

## Figures and Tables

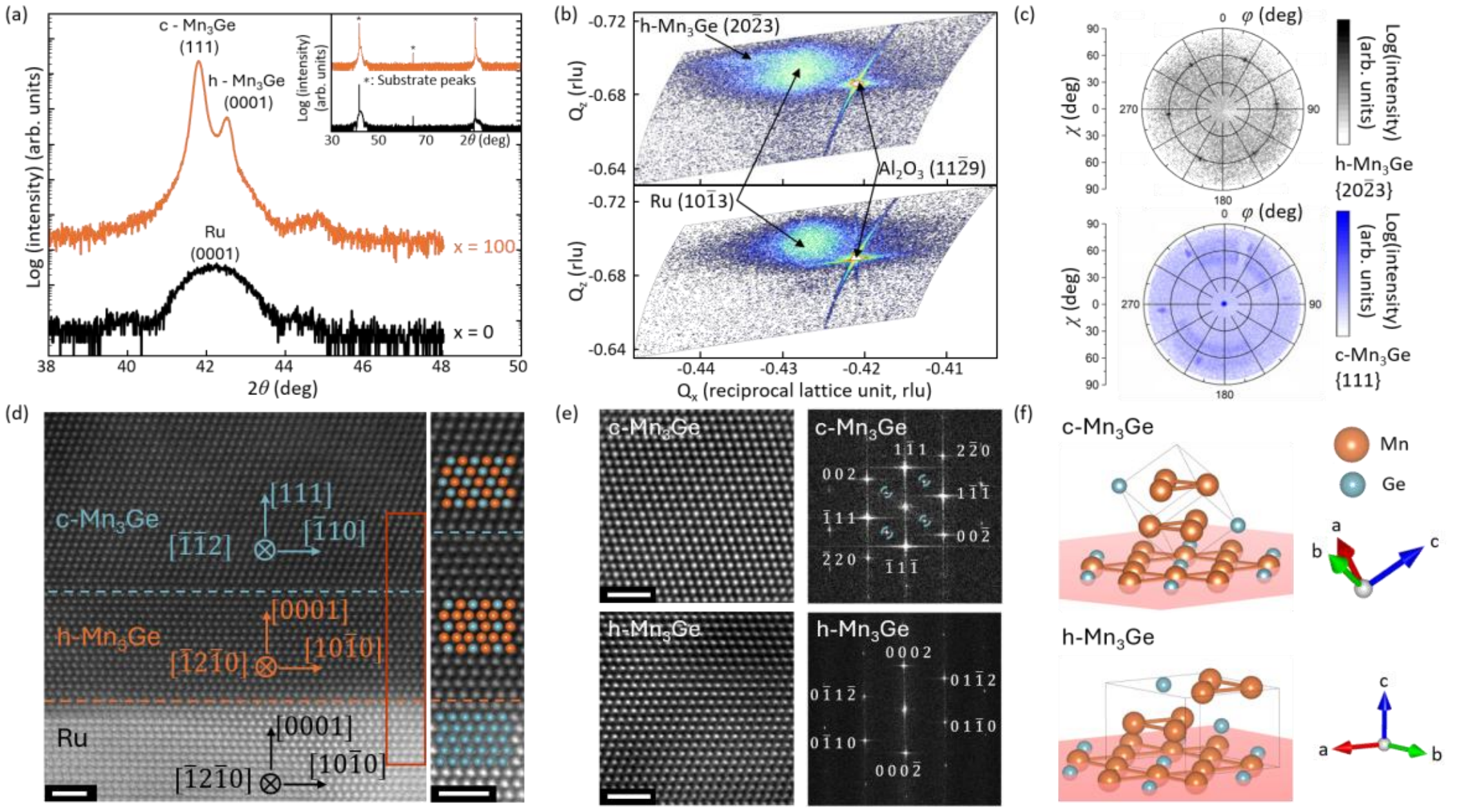

**Figure 1.** Structural characterization of $Mn_3Ge$ films. (a) $2\theta$-$\omega$ XRD scans with a 0.2° $\omega$ offset of $Mn_3Ge$ ($x$ nm)/Ru (5 nm)/$Al_2O_3$ films labeled with the structural phases. The inset shows scans without $\omega$ offset in a larger $2\theta$-$\omega$ range. (b) RSMs of a $Mn_3Ge$ (100 nm)/Ru (5 nm)/$Al_2O_3$ film (top) and a Ru (5 nm)/$Al_2O_3$ film (bottom), labeled with reflections of h-$Mn_3Ge$ $(20\bar{2}3)$, Ru $(10\bar{1}3)$, and $Al_2O_3$$(11\bar{2}9)$ planes. (c) $\chi$-$\phi$ pole figures of a $Mn_3Ge$ (100 nm)/Ru (5 nm)/$Al_2O_3$ film aligning on h-$Mn_3Ge$ $(20\bar{2}3)$ (top) and c-$Mn_3Ge$ (111) (bottom) reflections. (d) Cross-sectional STEM-HADDF image of a $Mn_3Ge$ (100 nm)/Ru (5 nm)/$Al_2O_3$ film, viewed along the $[1\bar{1}00]$ zone axis of the $Al_2O_3$ substrate. The boxed region is enlarged at right and matched with atomic models. The scale bars represent 1 nm. (e) STEM-HADDF images (left) and corresponding FFT patterns (right) for h-$Mn_3Ge$ and c-$Mn_3Ge$ regions. Reflections in the FFT patterns are indexed, and the superlattice reflections of c-$Mn_3Ge$ are indicated in the top-right panel. Scale bars represent 1 nm. (f) Atomic models of c-$Mn_3Ge$ (top) and h-$Mn_3Ge$ (bottom).

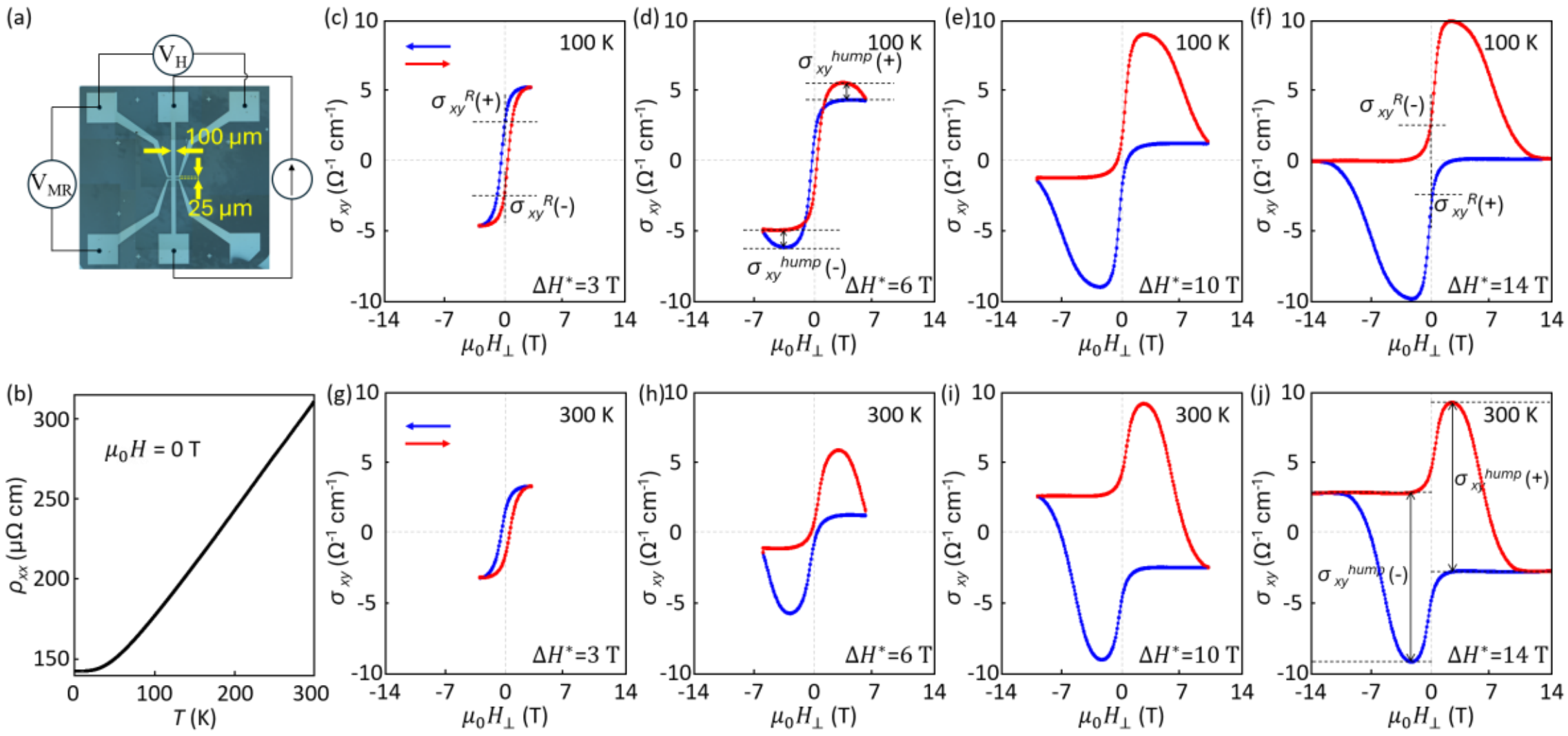

**Figure 2.** Magnetotransport measurements from a 100-nm-thick $Mn_3Ge$ film. (a) The Hall-bar geometry used in the measurements, with Hall-bar dimensions labelled. (b) Temperature dependent longitudinal resistivity. (c)-(j) $\sigma_{xy}$ hysteresis loops at 100 K (c)-(g) and 300 K (g)-(j), measured over $\Delta H^*$ = 3, 6, 10, and 14 T. The blue (red) line corresponds to the field sweep from positive to negative (negative to positive).

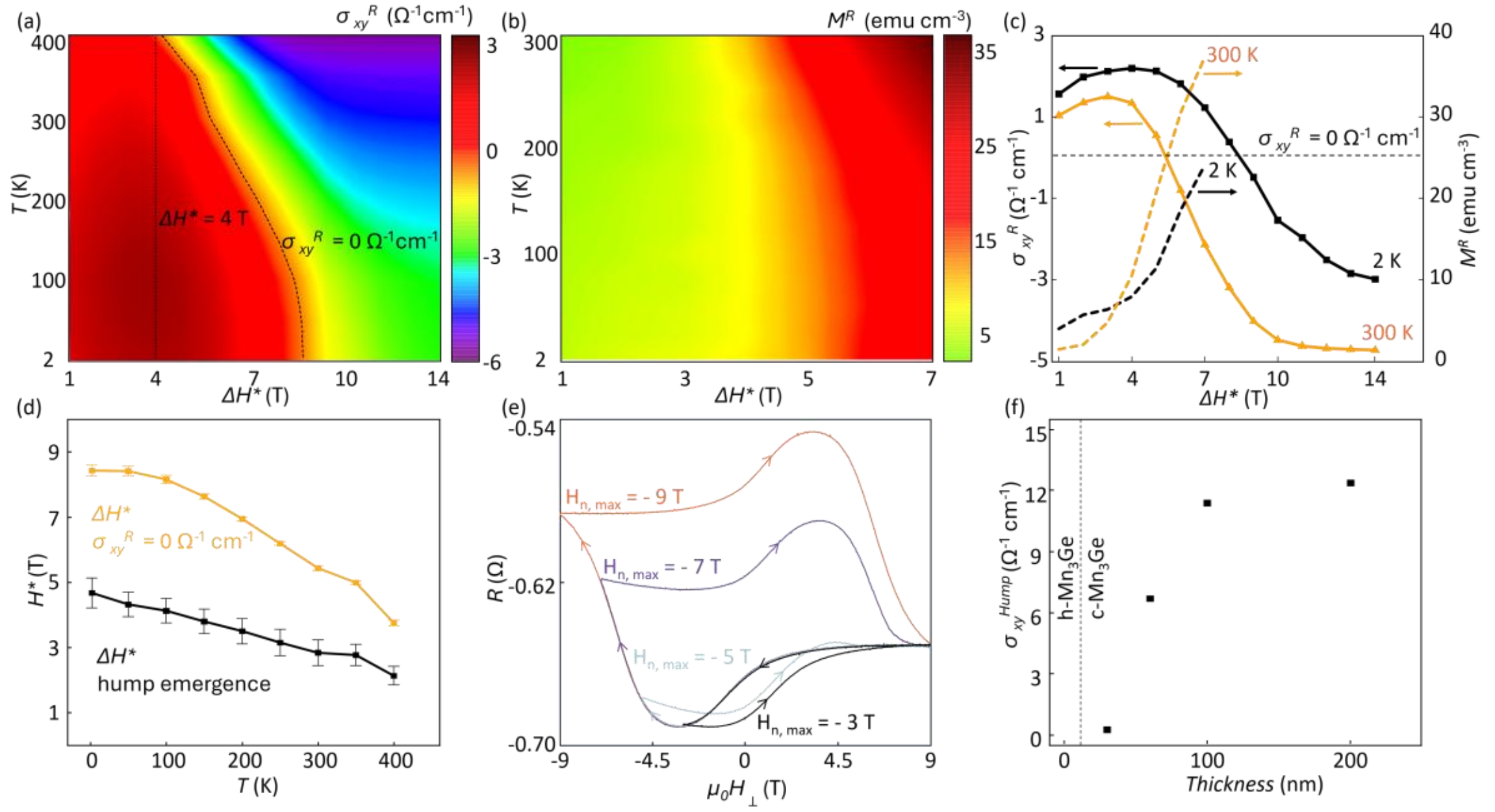

**Figure 3.** Magnetotransport properties of $Mn_3Ge$ films. (a) Color map of $\sigma_{xy}^R$ as a function of $\Delta H^*$ (1-14 T) and temperature (2-400 K). Dashed lines indicate where $\Delta H^*$ = 4 T and where $\sigma_{xy}^R = 0$. (b) Color map of $M^R$ as a function of $\Delta H^*$ (1-7 T) and temperature (2-300 K). (c) $M^R$ (dashed lines) $\sigma_{xy}^R$ (solid lines) as functions of $\Delta H^*$ at 2 K and 300 K. The horizontal dashed line indicates where $\sigma_{xy}^R = 0$. (d) $\Delta H^*$ at which the hump emerges and at which the sign of $\sigma_{xy}^R$ reverses as functions of temperature. (e) Minor hysteresis loops of $\sigma_{xy}$ from a 100-nm-thick $Mn_3Ge$ film at 300 K, labeled with the maximum negative field $H_{n,max}$. (f) Hump magnitude $\sigma_{xy}^{hump}$ as a function of film thickness at 300 K measured over $\Delta H^*$ = 9 T. The vertical dashed line indicates the hexagonal-cubic interface region.

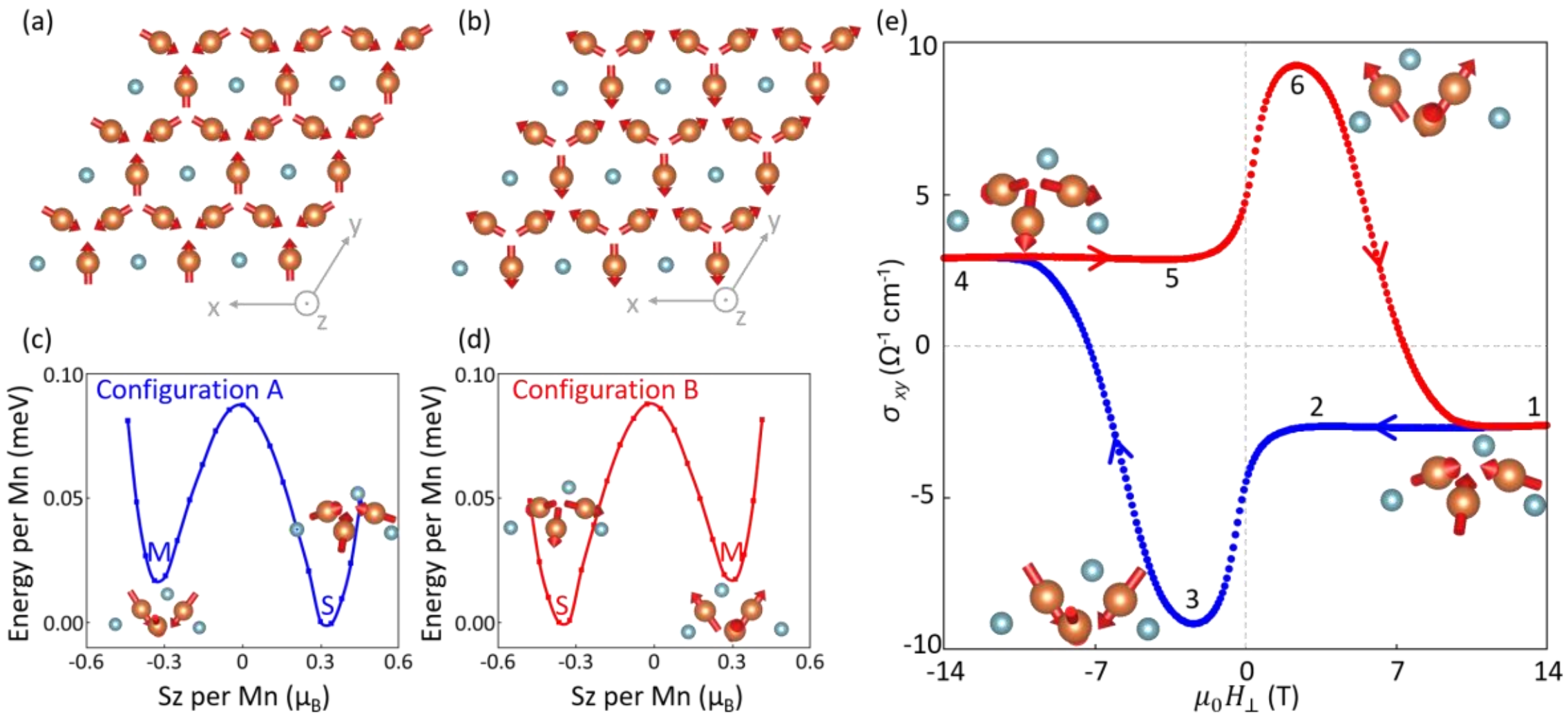

**Figure 4.** Spin configurations and magnetic states in c-$Mn_3Ge$. Spin configurations A (a) and B (b) in c-$Mn_3Ge$. The large and small spheres represent Mn and Ge atoms, respectively, and the red arrows represent the spin directions. The *x*, *y*, and *z* axes correspond to the $[11\bar{2}]$, $[\bar{1}21]$, and $[111]$ axes in c-$Mn_3Ge$, respectively. Total energy as a function of the out-of-plane canting for spin configurations A (c) and B (d), labeled with the stable (S) and the metastable (M) states. (e) A $\sigma_{xy}$ hysteresis loop from a 100-nm-thick film at 300 K over a magnetic field sweep range of $\Delta H^*$ = 14 T, labeled with schematic magnetic regimes 1-6 used in the phenomenological model. The schematics in (c) and (d) illustrate the out-of-plane spin canting, and those in (e) indicate the dominant spin configuration contributing to the AHE in each regime.